\documentclass[iop,useAMS,usenatbib]{emulateapj}

\usepackage{graphicx}
\usepackage[percent]{overpic}
\usepackage{multirow}
\usepackage{color}
\usepackage{rotating}
\usepackage{amsmath}
\usepackage{mathtools}
\usepackage{physics}
\usepackage[percent]{overpic}

\newcommand{\msun}{${\rm M_{\sun}}$}

\def\ltsima{$\; \buildrel < \over \sim \;$}
\def\simlt{\lower.5ex\hbox{\ltsima}}
\def\gtsima{$\; \buildrel > \over \sim \;$}
\def\simgt{\lower.5ex\hbox{\gtsima}}
\def\kms{{\rm\,km\,s^{-1}}}

\def\kpc{{\rm\,kpc}}
\def\mpc{{\rm\,Mpc}}
\def\msun{{\rm\,M_\odot}}

\newcommand{\fmmm}[1]{\mbox{$#1$}}
\newcommand{\scnd}{\mbox{\fmmm{''}\hskip-0.3em .}}
\newcommand{\scnp}{\mbox{\fmmm{''}}}

\newcommand{\mcnp}{\mbox{\fmmm{'}}}

\def\AA{$\; \buildrel \circ \over {\rm A}$}

\def\deg{^\circ}
\def\s{\ifmmode \widetilde \else \~\fi}
\def\={\overline}

\def\spose#1{\hbox to 0pt{#1\hss}}

\def\lta{\mathrel{\spose{\lower 3pt\hbox{$\mathchar"218$}}
     \raise 2.0pt\hbox{$\mathchar"13C$}}}
\def\gta{\mathrel{\spose{\lower 3pt\hbox{$\mathchar"218$}}
     \raise 2.0pt\hbox{$\mathchar"13E$}}}
\def\Dt{\spose{\raise 1.5ex\hbox{\hskip3pt$\mathchar"201$}}}    % upper case
\def\dt{\spose{\raise 1.0ex\hbox{\hskip2pt$\mathchar"201$}}}    % lower case

\def\dotsfill{\leaders\hbox to 1em{\hss.\hss}\hfill}

\def\Gyr{{\rm\,Gyr}}

\def\Gaia{{\it Gaia}}
\def\Euclid{{\it Euclid}}

\def\ltsima{$\; \buildrel < \over \sim \;$}
\def\gtsima{$\; \buildrel > \over \sim \;$}
\def\lsim{\lower.5ex\hbox{\ltsima}}
\def\gsim{\lower.5ex\hbox{\gtsima}}
\def\lapp{\ifmmode\stackrel{<}{_{\sim}}\else$\stackrel{<}{_{\sim}}$\fi}
\def\gapp{\ifmmode\stackrel{>}{_{\sim}}\else$\stackrel{<}{_{\sim}}$\fi}

\shorttitle{The CFIS-\MakeLowercase{u} survey}
\shortauthors{Ibata et al.}

\begin{document}

\title{The Canada-France Imaging Survey:\\
First results from the \MakeLowercase{{\it u}}-band component}

\author{Rodrigo A. Ibata\altaffilmark{1}}
\author{Alan McConnachie\altaffilmark{2}}
\author{Jean-Charles Cuillandre\altaffilmark{3}}
\author{Nicholas Fantin\altaffilmark{2}}
\author{Misha Haywood\altaffilmark{4}}
\author{Nicolas F. Martin\altaffilmark{2}}
\author{Piere Bergeron\altaffilmark{5}}
\author{Volker Beckmann\altaffilmark{6}}
\author{Edouard Bernard\altaffilmark{7}}
\author{Piercarlo Bonifacio\altaffilmark{4}}
\author{Elisabetta Caffau\altaffilmark{4}}
\author{Raymond Carlberg\altaffilmark{8}}
\author{Patrick C\^ot\'e\altaffilmark{2}}
\author{R\'emi Cabanac\altaffilmark{9}}
\author{Scott Chapman\altaffilmark{10}}
\author{Pierre-Alain Duc\altaffilmark{1}}
\author{Florence Durret\altaffilmark{11}}
\author{Beno\^it Famaey\altaffilmark{1}}
\author{S\'ebastien Frabbro\altaffilmark{2}}
\author{Stephen Gwyn\altaffilmark{2}}
\author{Francois Hammer\altaffilmark{4}}
\author{Vanessa Hill\altaffilmark{7}}
\author{Michael J. Hudson\altaffilmark{12}}
\author{Ariane Lan\c con\altaffilmark{1}}
\author{Geraint Lewis\altaffilmark{13}}
\author{Khyati Malhan\altaffilmark{1}}
\author{Paola di Matteo\altaffilmark{4}}
\author{Henry McCracken\altaffilmark{14}}
\author{Simona Mei\altaffilmark{15,16,17}}
\author{Yannick Mellier\altaffilmark{11}}
\author{Julio Navarro\altaffilmark{18}}
\author{Sandrine Pires\altaffilmark{3}}
\author{Chris Pritchet\altaffilmark{18}}
\author{Celine Reyl\'e\altaffilmark{19}}
\author{Harvey Richer\altaffilmark{20}}
\author{Annie C. Robin\altaffilmark{19}}
\author{Rub\'en S\'anchez Jannsen\altaffilmark{21}}
\author{Marcin Sawicki\altaffilmark{22}}
\author{Douglas Scott\altaffilmark{20}}
\author{Vivien Scottez\altaffilmark{11}}
\author{Kristine Spekkens\altaffilmark{23}}
\author{Else Starkenburg\altaffilmark{24}}
\author{Guillaume Thomas\altaffilmark{1}}
\author{Kim Venn\altaffilmark{18}}

\altaffiltext{1}{Observatoire Astronomique, Universit\'e de Strasbourg, CNRS, 11, rue de l'Universit\'e, F-67000 Strasbourg, France; rodrigo.ibata@astro.unistra.fr}

\altaffiltext{2}{NRC Herzberg Institute of Astrophysics, 5071 West Saanich Road, Victoria, BC, V9E 2E7, Canada}

\altaffiltext{3}{CEA-Saclay/Observatoire de Paris}

\altaffiltext{4}{GEPI, Observatoire de Paris, PSL Research University, CNRS, Place Jules Janssen, 92190 Meudon, France}

\altaffiltext{5}{D\'epartement de Physique, Universit\'e de Montr\'eal,   C.P.~6128, Succ.~Centre-Ville, Montr\'eal, Qu\'ebec H3C 3J7, Canada}

\altaffiltext{6}{CNRS / IN2P3, 3 rue Michel Ange, 75794 Paris Cedex 16, France}

\altaffiltext{7}{Universit\'e C\^ote d'Azur, OCA, CNRS, Lagrange, France}

\altaffiltext{8}{Department of Astronomy and Astrophysics, University of Toronto, Toronto, ON M5S 3H4, Canada}

\altaffiltext{9}{Institut de Recherche en Astrophysique et Plan\'etologie, Observatoire Midi-Pyrenees, 65000 Tarbes, France}

\altaffiltext{10}{Department of Physics and Atmospheric Science, Dalhousie University, Coburg Road, Halifax, NS B3H 1A6, Canada}

\altaffiltext{11}{Institut d'Astrophysique de Paris, UMR 7095 CNRS, Universit\'e Pierre et Marie Curie, 98bis boulevard Arago, 75014, Paris, France}

\altaffiltext{12}{Dept. of Physics \& Astronomy, University of Waterloo, Waterloo, ON N2L 3G1 Canada}

\altaffiltext{13}{Sydney Institute for Astronomy, School of Physics, A28, University of Sydney, NSW 2006, Australia}

\altaffiltext{14}{Sorbonne Universit\'e, UPMC Univ Paris 06, UMR 7095, Institut d'Astrophysique de Paris, F-75014, Paris, France}

\altaffiltext{15}{LERMA, Observatoire de Paris,  PSL Research University, CNRS, Sorbonne Universit\'es, UPMC Univ. Paris 06, F-75014 Paris, France}

\altaffiltext{16}{University of Paris Denis Diderot, University of Paris Sorbonne Cit\'e (PSC), 75205 Paris Cedex 13, France}

\altaffiltext{17}{Jet Propulsion Laboratory, Cahill Center for Astronomy \& Astrophysics, California Institute of Technology, 4800 Oak Grove Drive, Pasadena, California, USA}

\altaffiltext{18}{Department of Physics and Astronomy, University of Victoria, Victoria, BC, V8P 1A1, Canada}

\altaffiltext{19}{Institut UTINAM, CNRS UMR6213, Univ. Bourgogne Franche-Comt\'e, OSU THETA Franche-Comt\'e-Bourgogne, Observatoire de Besançon, BP 1615, 25010 Besan{\c c}on Cedex, France}

\altaffiltext{20}{Dept. of Physics and Astronomy, University of British Columbia, Vancouver, B.C. V6T 1Z1, Canada}

\altaffiltext{21}{STFC UK Astronomy Technology Centre, The Royal Observatory Edinburgh, Blackford Hill, Edinburgh, EH9 3HJ, UK}

\altaffiltext{22}{Department of Astronomy and Physics and Institute for Computational Astrophysics, Saint Mary's University, 923 Robie Street, Halifax, Nova Scotia B3H 3C3, Canada}

\altaffiltext{23}{Department of Physics, Royal Military College of Canada, PO Box 17000, Station Forces, Kingston, Ontario, Canada}

\altaffiltext{24}{Leibniz Institute for Astrophysics Potsdam (AIP), An der Sternwarte 16, D-14482 Potsdam, Germany}

\begin{abstract}
The Canada-France Imaging Survey (CFIS) will map the northern high Galactic latitude sky in the $u$-band (``CFIS-u", 10,000$\, {\rm deg^2}$) and in the $r$-band (``CFIS-r", 5,000$\, {\rm deg^2}$), enabling a host of stand-alone science investigations, and providing some of the ground-based data necessary for photometric redshift determination for the \Euclid\ mission. In this first contribution we present the $u$-band component of the survey, describe the observational strategy, and discuss some first highlight results, based on approximately one third of the final area. We show that the Galactic anticenter structure is distributed continuously along the line of sight, out to beyond $20\kpc$, and possesses a metallicity distribution that is essentially identical to that of the outer disk sampled by APOGEE. This suggests that it is probably a buckled disk of old metal-rich stars, rather than a stream or a flare. We also discuss the future potential for CFIS-u in discovering star-forming dwarf galaxies around the Local Group, the characterization of the white dwarf and blue straggler population of the Milky Way, as well as its sensitivity to low-surface brightness structures in external galaxies.
\end{abstract}


\section{Introduction}
\label{sec:Introduction}

Progress in astrophysics has depended, to a great extent, on the large imaging surveys that have provided the community with sources for study. From the Schmidt telescope campaigns of the 1950s to 1990s, to the digital era with the Sloan Digital Sky Survey (SDSS, \citealt{2000AJ....120.1579Y}) and Pan-STARRS1 (PS1, \citealt{2016arXiv161205560C}), almost every area of astrophysics has been influenced by the rich trove of morphological and photometric information large optical imaging surveys have provided.

Over the coming years, several new important sky surveys will undoubtedly continue this trend. These include the Large Synoptic Survey Telescope (LSST, \citealt{Collaboration:2012uk}) which will explore the time domain by repeatedly imaging approximately half the sky in six bands over the course of a decade; the \Gaia\ space mission \citep{2016A&A...595A...2G} which will unveil the astrometric sky; and the \Euclid\ satellite \citep{2011arXiv1110.3193L,2016SPIE.9904E..0OR}, which will gather data of superb image quality over 15,000$\, {\rm deg^2}$ to probe the dark universe via gravitational lensing. Mining and understanding this massive new parameter space will undoubtedly provide the basis of much of our scientific discussion for decades to come.

In the optical bands, the most challenging component of a sky survey to acquire is the $u$-band (centered around $3500$\AA). The difficulty comes from the intrinsic faintness of most sources, whose spectral energy distributions peak red-wards of ${u}$, it also stems from the poor transparency of the atmosphere in this band (together with the strong variation with airmass), and from the relatively poor efficiency of most optical CCD cameras in this wavelength region. This typically means that $u$-band observations are more expensive to obtain in telescope time than longer wavelength optical bands. It is the reason that the SDSS $u$-band is relatively shallow compared to the SDSS ${g,r,i}$-bands, and it probably accounts for the lack of a $u$-band in the PS1 survey.

Nevertheless, the $u$-band contains very important astrophysical information: in stars, the many metal lines in this spectral region render the $u$-band very useful for measuring metallicity; young or hot stellar populations have their greatest contrast in the UV, making this an important band to study star-formation in the nearby Universe; while for distant galaxies the $u$-band is very powerful for helping us to distinguish between photometric redshift solutions. 

These are some of the reasons why undertaking an all-sky $u$-band survey is of prime astrophysical importance. These reasons undoubtedly also motivated the {\it SkyMapper} \citep{2007PASA...24....1K} and {\it SCUSS} \citep{Zou:2016gx} projects. {\it SkyMapper} is currently surveying the whole southern sky to $u \sim 20$ at ${\rm S/N}=30$, while {\it SCUSS} has mapped the Southern Galactic Cap, reaching $u \sim 23$ at ${\rm S/N}=5$. Over the course of the next decade, LSST will also survey the southern sky in the $u$-band, but plans for a deep northern sky $u$-band survey have been missing, until now. One of the main motivations of our community in undertaking a northern deep $u$-band survey was to contribute to the photometric redshift measurements that are needed for the \Euclid\ mission. Ground-based photometric redshifts are essential for \Euclid, since the gravitational lensing and baryon acoustic oscillation analyses depend on distance, for which redshift can be used as a proxy.\footnote{The core science of \Euclid\ requires photometric redshifts derived from $griz$ ground-based photometry, plus the wide optical (``VIS") and near infrared ($JHK$) photometry measured on-board the satellite. Ideally, \Euclid\ would have access to $u$-band photometry over its entire 15,000$\, {\rm deg^2}$ survey area to $u=24.2$ (${\rm S/N}=10$, integrated over $2\scnp$ diameter apertures). Although CFIS-u only reaches $u=23.7$ with this metric, it will nevertheless be very helpful for rejecting nearer and more compact sources. It is expected that LSST will provide the required ground-based photometry in the South.}

However, the immediate scientific driver for the present $u$-band survey is Galactic Archaeology, in particular by using metallicity as a population discriminant, as well as as a means to improve photometric distance measurements for main sequence stars. The sensitivity of the $u$-band to metallicity was beautifully demonstrated by \citet[][hereafter I08]{2008ApJ...684..287I}, who undertook an analysis of SDSS photometry that allowed them to map out the metallicity structure of the Milky Way within about $9\kpc$ of the Sun. The limiting distance was set by the photometric depth of the SDSS $u$-band data, which substantially limited the volume that I08 could study in this way. As we discuss in the companion paper to this contribution (Ibata et al. 2017; hereafter, Paper~II), for the purpose of measuring photometric metallicities, the SDSS is effectively 2.7~mag too shallow in the $u$-band to make optimal use of its $g$-band depth. Providing this missing information is one of the aims of our survey, and we argue in paper~II that it is essential to study the Galactic halo in combination with \Gaia. The reason for this is that most halo stars in \Gaia\ will have faint magnitudes, beyond the threshold for accurate \Gaia\ parallax or metallicity measurements. By providing a reliable photometric metallicity and thereby a good photometric distance, it will be possible to convert \Gaia's proper motions (which remain excellent for all of their surveyed stars) into transverse velocities, which are physically much more useful.

The outline of this paper is as follows. In Section~\ref{sec:Observations} we explain the survey strategy, observations and data reduction. In Section~\ref{sec:Anticenter} we explore the stellar populations and structure towards the Galactic Anticenter, and we briefly discuss Galactic white dwarfs in Section~\ref{sec:WDs}. We show how CFIS-u can be used to uncover nearby star-forming dwarf galaxies in Section~\ref{sec:star-formation}, and discuss its sensitivity to low surface brightness structures in Section~\ref{sec:LSB}. We draw conclusions for our study in Section~\ref{sec:Conclusions}.

\section{CFIS-u}
\label{sec:Observations}

\begin{figure*}
\begin{center}
\begin{overpic}[angle=0, viewport= 60 310 790 525, clip, width=\hsize]{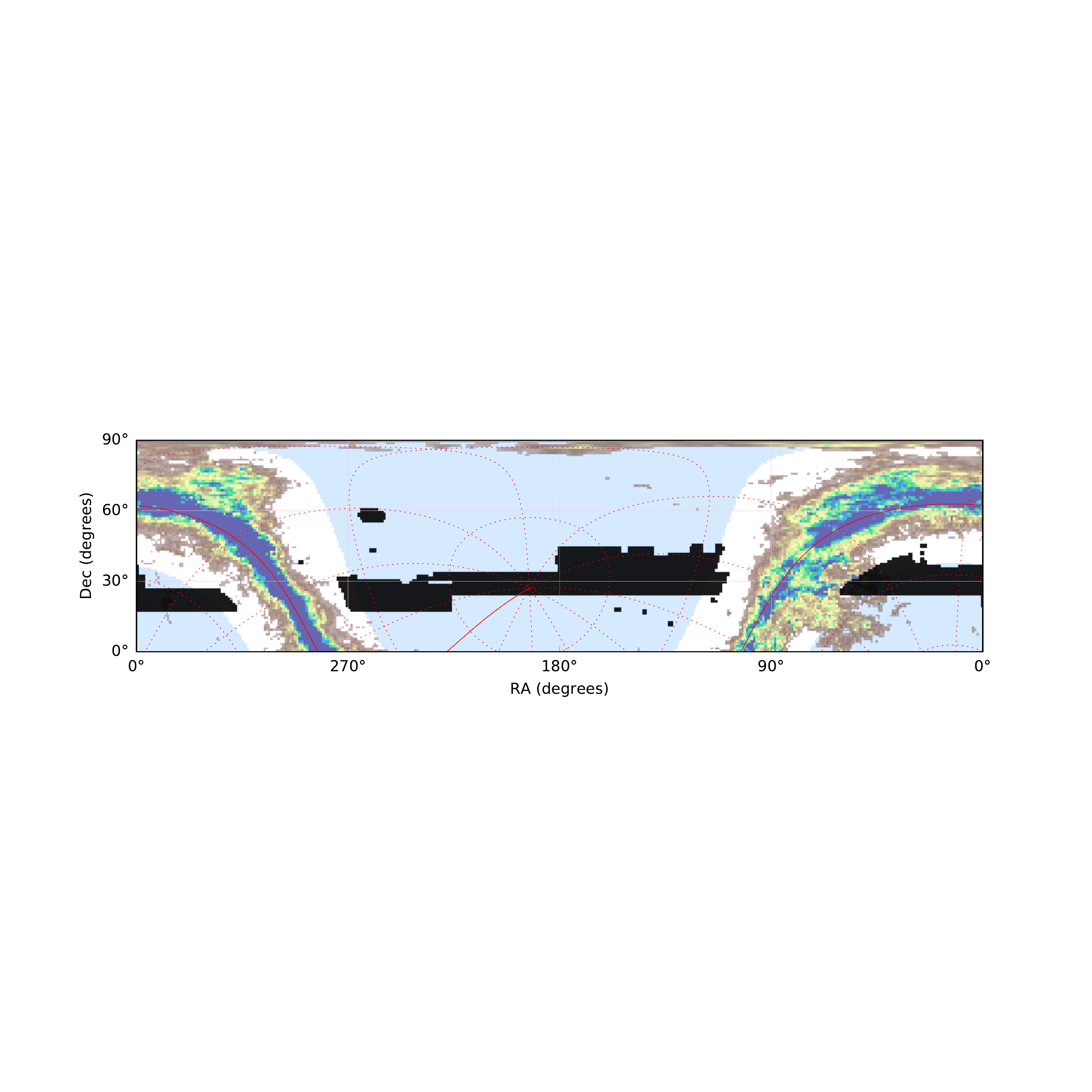}
\put(7,7){\includegraphics[, viewport= 201 253 392 491, clip, width=3cm]{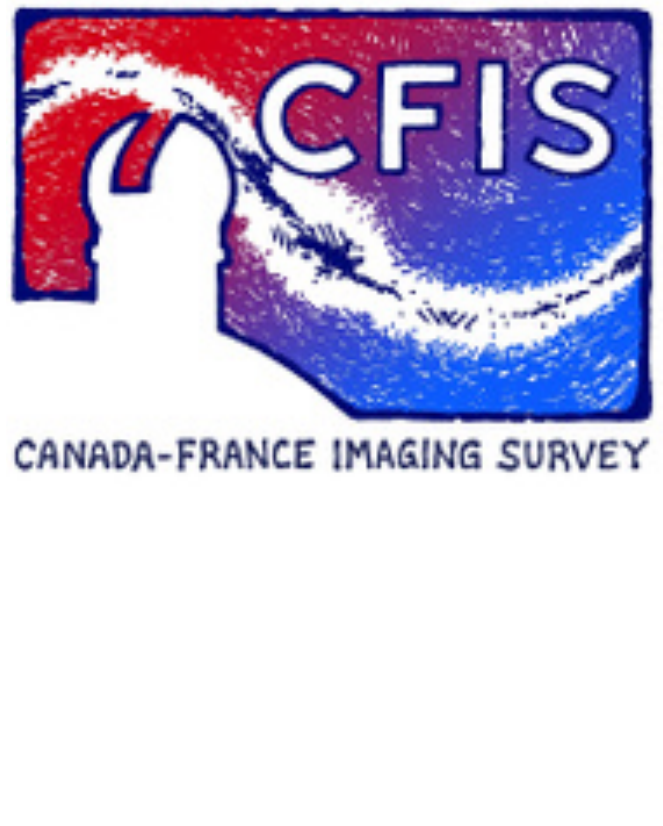}}
\end{overpic}
%\includegraphics[angle=0, viewport= 60 310 790 525, clip, width=\hsize]{fig01.pdf}
%\Put(10,50){\color{blue}\includegraphics[height=3cm]{CFIS-LogoWithFullName-Small.jpg}}\pause
\end{center}
\caption{Current footprint of the CFIS-u survey (solid black areas) on an equatorial projection of the northern sky. The light blue background shows the expected final footprint of the survey ($|b| > 25\deg$), once it is completed in 2020. The colormap that increases towards the Galactic plane shows the interstellar reddening, clipped so that the lowest value (in brown) marks $E(B-V)=0.15$. The red lines show Galactic coordinates, with the Galactic Plane and Galactic minor axis highlighted with a solid line. The isolated black ``islands'' that are disconnected from the main survey region are calibration targets, containing star clusters and the Draco dSph.}
\label{fig:footprint}
\end{figure*}

It was for the afore-mentioned reasons that we decided in late 2013 to propose a large $u$-band survey in response to a call for Large Programs at the Canada-France-Hawaii Telescope (CFHT) for the four semesters 2015A--2016B. This first survey was named the ``CFHT Legacy for the $u$-band All-Sky Universe'' (Luau) survey, and was awarded an initial 63.6 nights over this period to survey $\sim$3,500$\, {\rm deg^2}$. This program was designed to take advantage of the unique $u$-band sensitivity of CFHT with respect to all other current, large, ground-based facilities. In a subsequent call for the 2017--2019 period, we extended the scope of this wide-field survey to include extensive $r$-band, as well as $u$-band, imaging. The new program is called the Canada-France Imaging Survey, and consists of 271 nights of new data, in addition to the original allocation for Luau. The completed CFIS-u component (which includes the original Luau survey) will cover 10,000$\, {\rm deg^2}$, and the new $r$-band component will cover 5,000$\, {\rm deg^2}$. CFIS is intended to enable a broad swath of stand-alone science investigations, to contribute some of the necessary data for the derivation of photometric redshifts for the \Euclid\ space mission, and to provide a high quality legacy dataset for the northern hemisphere sky. More details on CFIS, in particular with relation to the $r$-band component, will be presented in a forthcoming contribution. Here, we focus on the data acquisition and processing and some science highlights using the existing $u$-band data.

One of the primary aims of CFIS-u is to improve upon I08 by providing much deeper $u$-band data that can be used to reach the full depth of the SDSS and PS1 surveys. The plan is for CFIS-u to cover the entire northern sky not optimally accessible to LSST in the $u$-band, and away from the Galactic plane $(|b|>25\deg)$, to a depth of ${u=24.4}$ with a photometric uncertainty of $\delta {u} \sim 0.2$~mag. The survey footprint is shown in Figure~\ref{fig:footprint}. In combination with $grizy$ data from SDSS and PS1, this enables the application of the techniques pioneered by I08 to a volume a factor of about $100$ greater than what can be reached with existing SDSS $u$-band data.

\begin{figure}
\begin{center}
\includegraphics[angle=0, viewport= 5 20 425 420, clip, width=\hsize]{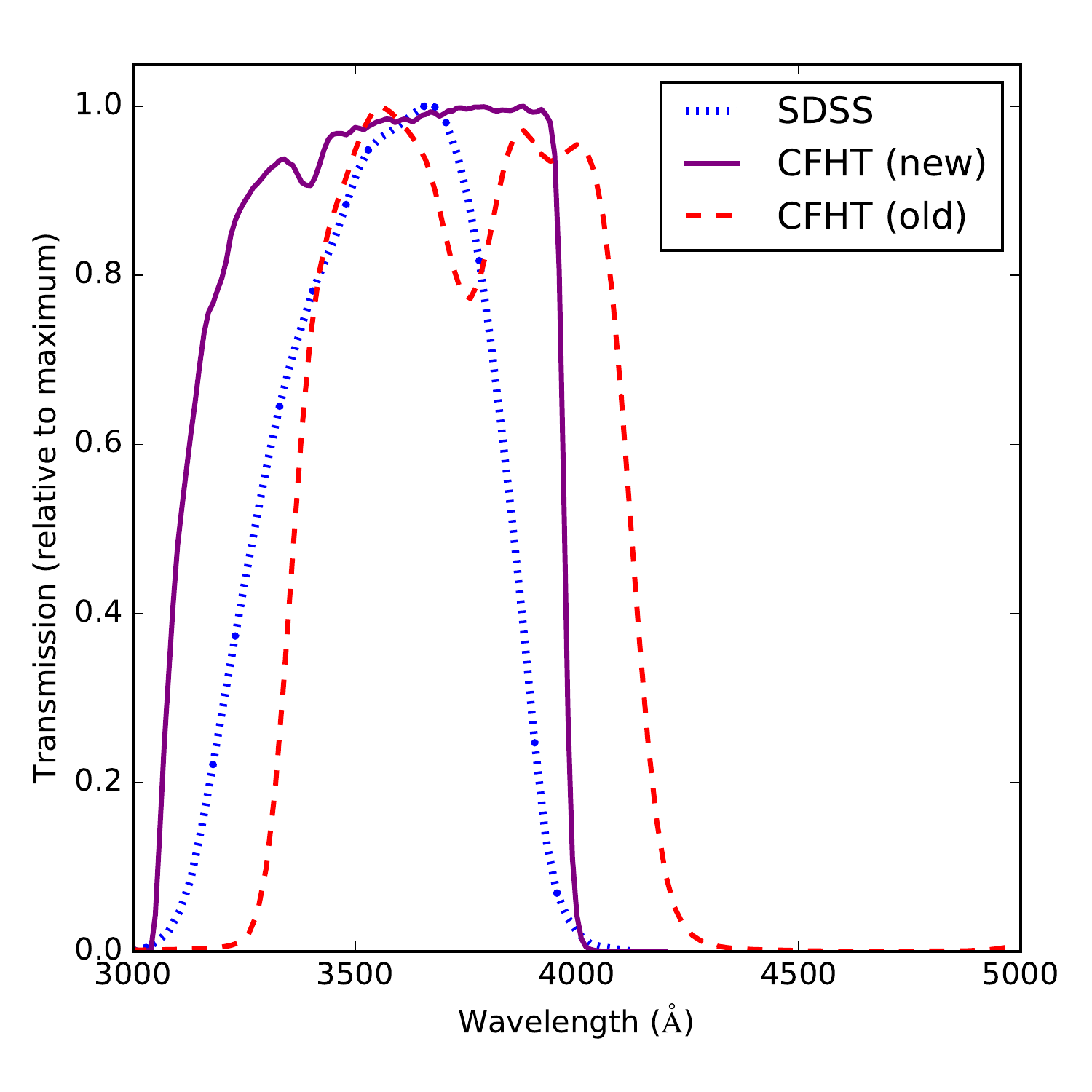}
\end{center}
\caption{Transmission curves for the SDSS and both old and new CFHT $u$-band filters. The new CFHT $u$-band (solid purple line) remains highly transmissive into the red ($>50$\% transmission until $\sim 3970$\AA), whereas the SDSS $u$-band (blue dotted line) drops rapidly after $\sim 3820$\AA. Thus the difference between the SDSS and new CFHT filters is sensitive to the \ion{Ca}{2} H$+$K lines ($3968.5$\AA\ and $3933.7$\AA) in the survey stars, among other features.}
\label{fig:filters}
\end{figure}

The imaging data for CFIS-u presented herein were obtained at the Canada-France Hawaii Telescope in semesters 2015A-2016B and are scheduled to continue until the end of 2019B (Jan 2020). The survey uses the MegaCam wide-field camera \citep{2003SPIE.4841...72B} with a new $u$-band filter procured in late 2014 which is physically wider than the $u$-band filter previously installed, allowing for the illumination of the full 40 CCDs of the MegaCam mosaic (previous filters only allowed for 36 CCDs to be illuminated). The new $u$-band filter is also bluer than the old one, being closer to the SDSS $u$-band, but with a significantly different transmission curve (see Figure~\ref{fig:filters}).

\begin{figure*}
\begin{center}
\includegraphics[angle=0, viewport= 45 5 780 335, clip, width=\hsize]{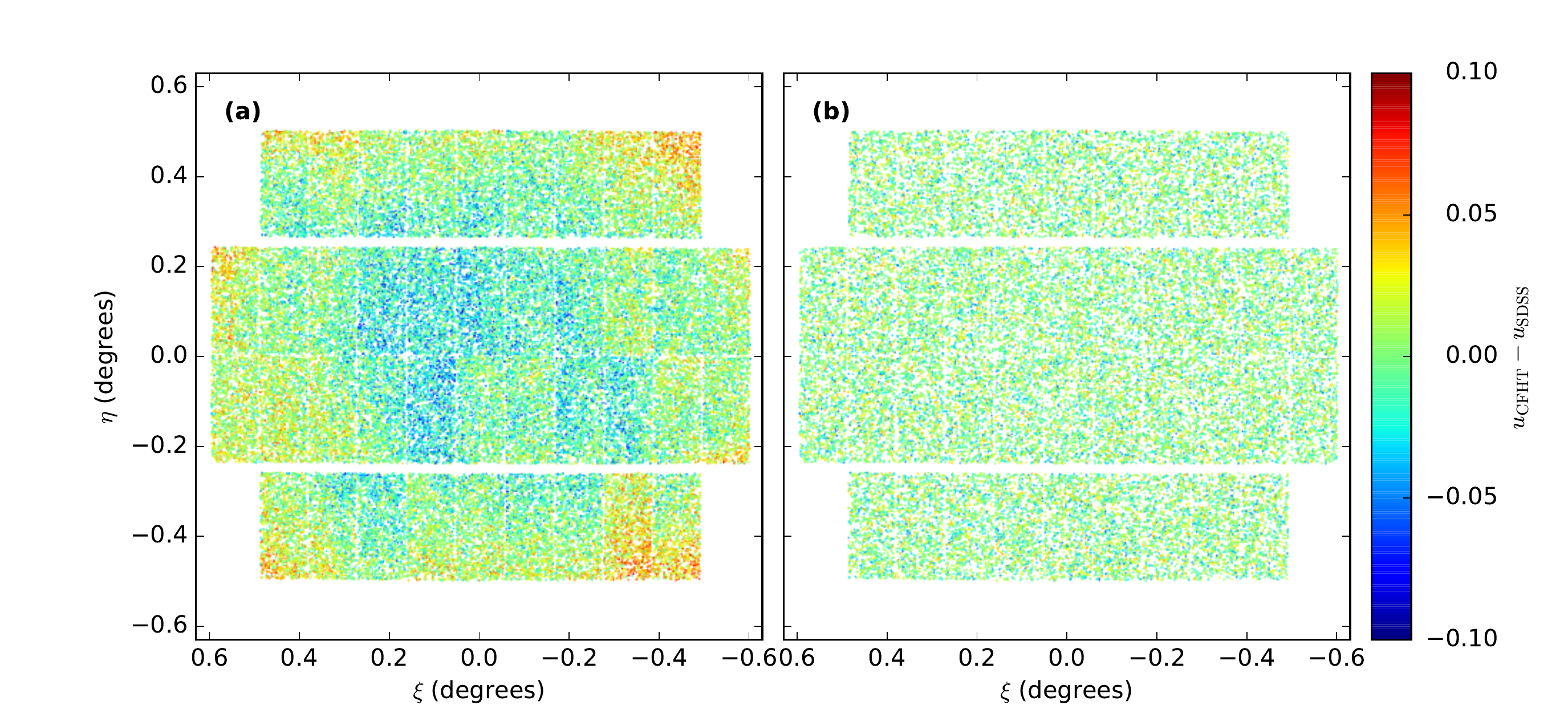}
\end{center}
\caption{Map of the photometric differences between the CFHT and SDSS DR13 measurements, plotted in standard (gnomonic) projection with respect to the center of the CFHT field of view. The data shown are all bright stars (${16<u_0<18}$) taken on the first run of the survey (run ``15Am04''). (a) shows the raw differences, whereas (b) shows these differences after correction using the two-dimensional Legendre polynomial model described in the text. While the raw photometry displays large systematic differences exceeding $\pm0.05$~mag, the corrected photometry is flat to better than 0.02~mag rms. The data on other runs show a very similar distribution of photometric differences.}
\label{fig:phot_diff}
\end{figure*}

We adopted a very simple tiling pattern, using the central rectangular ($9\times4$ CCD) regions butted against each other in the East-West direction with only a small overlap that was optimised over the course of the survey. The four lateral CCDs (of which there are two on each of the Eastern and Western ends of the camera -- see Figure~\ref{fig:phot_diff}) then overlap 100\% in the adjacent fields, both East and West.

As of 2017 February 25th, a total of 10,827 useful images have been processed for this large survey, covering an area of approximately 2,909$\, {\rm deg^2}$ of sky, shown in black in Figure~\ref{fig:footprint}. The median seeing over this set of images is $0\scnd78$. The strategy we adopted was to take exposures of 80~s per field, but to stagger the tiling pattern so that successive East-West rows are displaced from the previous row by one third of a field both in the North and East directions. Thus the final exposure depth is 240~s over most of the survey. However, there are small horizontal gaps ($80\scnp$) in the MegaCam camera between CCD rows, and in those areas the coadded exposure is only 160~s. In about $10$\% of the survey the overlap with the four lateral CCDs gives 320~s of total exposure. 

The images were initially pre-processed (de-biassed and flat-fielded) at the CFHT using the {\tt Elixir} pipeline software \citep{2004PASP..116..449M}. We then used the Cambridge Astronomical Survey Unit (CASU) photometric pipeline \citep{Irwin:2001eq} to detect sources and derive aperture magnitudes for all sources $1.5\sigma$ above the local background in each individual frame (i.e. not stacked). An aperture radius of four pixels was used ($0\scnd75$). In this process, the astrometry was calibrated with respect to the \Gaia\ Data Release 1 (DR1) catalog \citep{2016A&A...595A...2G}, although in a small fraction of frames ($\ll 1$\%) we were forced to use PS1 positions due to a paucity of \Gaia\ stars in these high-latitude fields. The typical rms scatter in the astrometric solutions is excellent, being $0\scnd034$.

The source classification was undertaken using a customized version of the CASU image classification tool, which examines the morphology of individual sources. The main improvement we implemented to this algorithm was to train the point-source criteria in each field using \Gaia\ stars (which allows better rejection of extended sources and noise).

To compare the CFHT photometry to the SDSS DR13 photometry \citep{Albareti:2016tq}, we made use of the following color transformation (used for calibration at the Canadian Astronomy Data Centre):
\begin{equation}
{u_{\rm CFHT} = u_{\rm SDSS} + 0.038 (u_{\rm SDSS} - g_{\rm SDSS}) - 0.165} \, ,
\end{equation}
which is found to be valid for ${u_{\rm SDSS} - g_{\rm SDSS} > 1.3}$ (note that this transformation is only valid for stars that are redder than the main sequence stars studied in Paper~II). After comparing our photometry to the SDSS $u$-band (transformed with this color equation), we found that there was a strong pattern of residuals over the camera. This problem has been noted before (e.g., \citealt{Regnault:2009bk,2014ApJ...780..128I}), and is due to the CFHT pipeline creating slightly incorrect flats. The master flats are created by the pipeline from all the twilight observations during a run (MegaCam is mounted on the CFHT typically once per month for a two-week period, or ``run''), but this introduces a flat-fielding error with respect to nighttime science images. (In principle, it should be possible to use the dark-sky CFIS-u data themselves to generate a better $u$-band flat, but we did not attempt to implement this). The SDSS-MegaCam pattern for the $u$-band appears to have remained stable over the course of the survey, but since we have sufficiently good statistics, we decided to model it on a run-by-run basis. For each CCD the residual pattern was modeled as a two-dimensional Legendre polynomial with up to cubic terms in $x$ and $y$ (i.e., 10 parameters in all), and the resulting models were used to flatten all the data. An example of this correction for one of the runs is shown in Figure~\ref{fig:phot_diff}. The typical rms scatter for stars in the range $16< {u} < 19$ (and ${1.3 < u-g < 2.5}$) is $0.02$~mags. Note that even with perfectly calibrated CFHT and SDSS photometry, there would be scatter in the photometric differences due to the fact that the two $u$-band filters are significantly different (see Figure~\ref{fig:filters}). 

\begin{figure*}
\begin{center}
\hbox{
\includegraphics[angle=0, viewport= 70 5 660 405, clip, height=5.4cm]{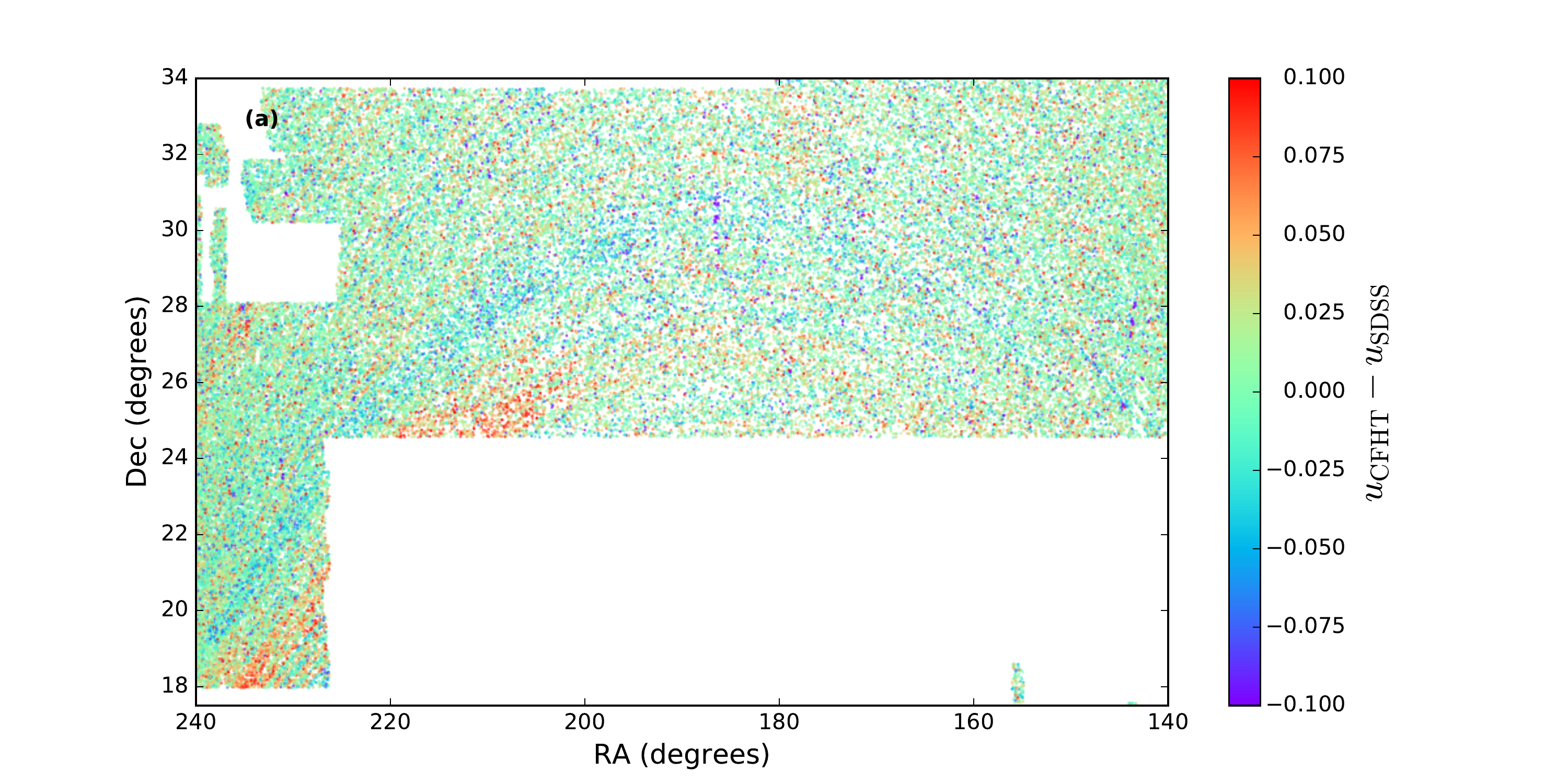}
\includegraphics[angle=0, viewport= 50 5 770 405, clip, height=5.4cm]{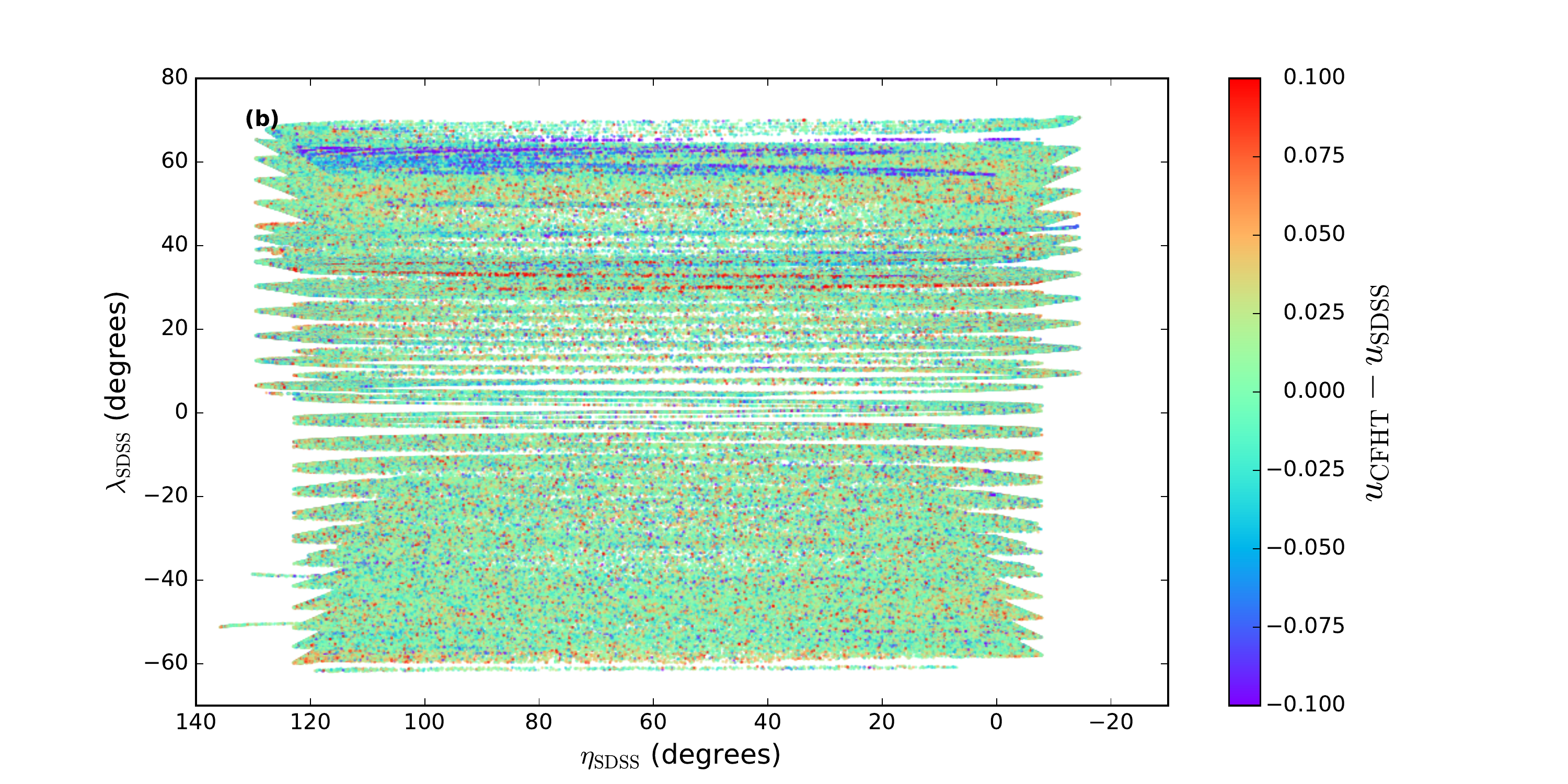}
}
\end{center}
\caption{(a) Zoomed-in sky map of the photometric differences between the CFHT and SDSS DR13 measurements. The objects that are plotted are high signal to noise stars with ${16<u_0<19}$ within the color range ${1.3 < u_{\rm SDSS,0}-g_{\rm SDSS,0} < 1.7}$. The striking stripy pattern that is visible all over this region (and elsewhere in the survey) follows precisely the SDSS ``stripes''. (b) This behaviour is even more obvious when the CFIS-u stars are plotted in SDSS $\eta,\lambda$ coordinates, which is a coordinate system in which the SDSS stripes are arranged horizontally. This clearly shows that the SDSS has substantial position-dependent errors in the $u$-band. In contrast, in panel (a) one does not see any pattern of differences that could be obviously associated with the CFHT (which would cause a roughly $1\deg \times 1\deg$ pattern aligned with the cardinal directions).}
\label{fig:phot_diff_SDSS}
\end{figure*}

\begin{figure}
\begin{center}
\includegraphics[angle=0, viewport= 1 13 400 400, clip, width=\hsize]{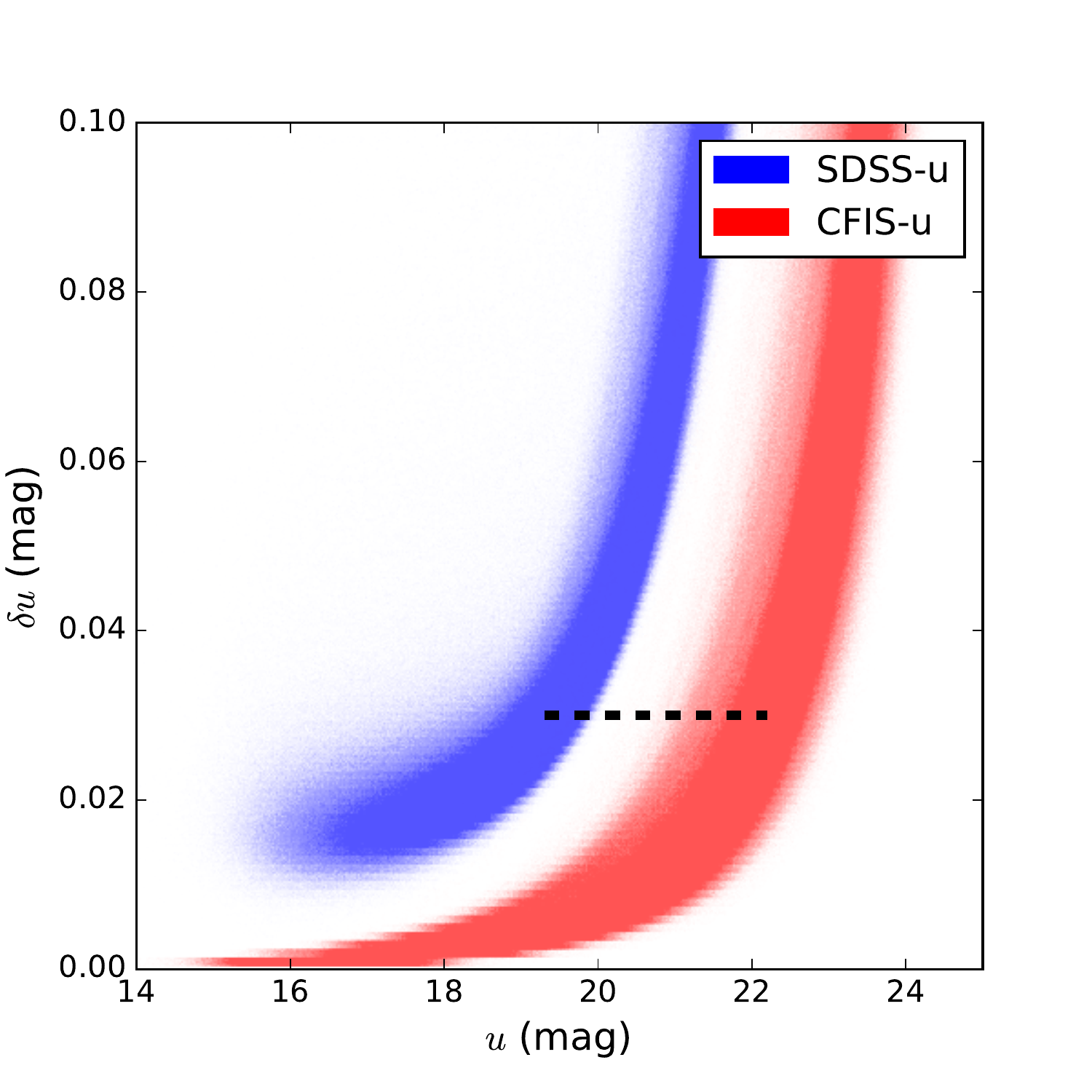}
\end{center}
\caption{Distribution of uncertainties as a function of magnitude for the SDSS (blue) and CFIS-u (red). At the limiting uncertainty for good metallicity measurements of $\delta {u}=0.03$, CFIS-u is approximately 3~mag deeper than the SDSS (black dashed line).}
\label{fig:uncertainties}
\end{figure}

The photometric offsets between overlapping fields, and with the SDSS, were then used to derive a global solution to the zero point of the survey. Due to the size of the matrix inversion problem ($>98,000$ equations with 10,827 unknown zero points), we simplified the task by solving for the zero points using data in bands of $30\deg$ in RA, but offset in $10\deg$ intervals. In this way we solve for the central $10\deg$, while maintaining the information of the photometric offsets $10\deg$ on either side of that region. Any equation solutions (i.e., field-to-field photometric differences) that were discrepant at more than $2.5\sigma$ were discarded, and the procedure was iterated (keeping the plausible field-to-field photometric offsets) until convergence. We checked that the zero points in the overlapping RA bands were consistent to better than $0.01~{\rm mag}$.

Finally, the individual sources between frames were matched using a $0\scnd3$ search radius, and their photometric measurements were combined using the photometric uncertainties to construct a weighted flux average, and hence a magnitude. This results in a catalog containing $>3.0\times 10^7$ sources. These detections were then matched against the SDSS, or PS1, using a $0\scnd5$ search radius, thus matching all but the very highest, rare, proper motion sources.

In comparing the corrected CFHT photometry with the SDSS DR13 data (which was re-calibrated by the SDSS team as detailed in \citealt{2016ApJ...822...66F}), we noticed that strong residuals remain. However, these residuals do not follow the MegaCam footprint, but quite obviously track the SDSS observing pattern of ``stripes'', as we show in Figure~\ref{fig:phot_diff_SDSS}. Once the CFIS-u survey is completed, we will be able to use it to fully recalibrate the SDSS $u$-band. In the meantime, some caution is needed when using the SDSS $u$, which clearly contains spatially-correlated zero point errors.

The photometric uncertainties in the CFHT data are derived from the square root of the variance of the weighted flux measurements. The distribution of these uncertainties is displayed in Figure~\ref{fig:uncertainties}, along with the uncertainties of the SDSS $u$-band photometry of the same targets. The CFHT photometry can be seen to be significantly deeper than the SDSS, typically reaching $\sim 3$~magnitudes fainter at the same uncertainty level.

Two additional photometric catalogs were generated, using the SDSS DR13 positions of point sources and all PS1 sources as input centroids for forced photometry on the CFIS $u$-band images. For the SDSS, it was necessary first to shift the astrometric solution of all CFIS frames onto the SDSS solution, which is slightly different to that of \Gaia\ (this is not necessary for PS1, as it is already on the \Gaia\ astrometric reference frame). As before, we measured $u$-band aperture magnitudes within a four pixel radius centered on all those source positions. The resulting forced photometry catalogs have (at present) $>2.0\times 10^7$ and $>1.4\times 10^8$ measurements, at the positions of the SDSS point-sources and PS1 detections, respectively.

\begin{figure*}
\begin{center}
\includegraphics[angle=0, viewport= 65 50 790 790, clip, width=15.0cm]{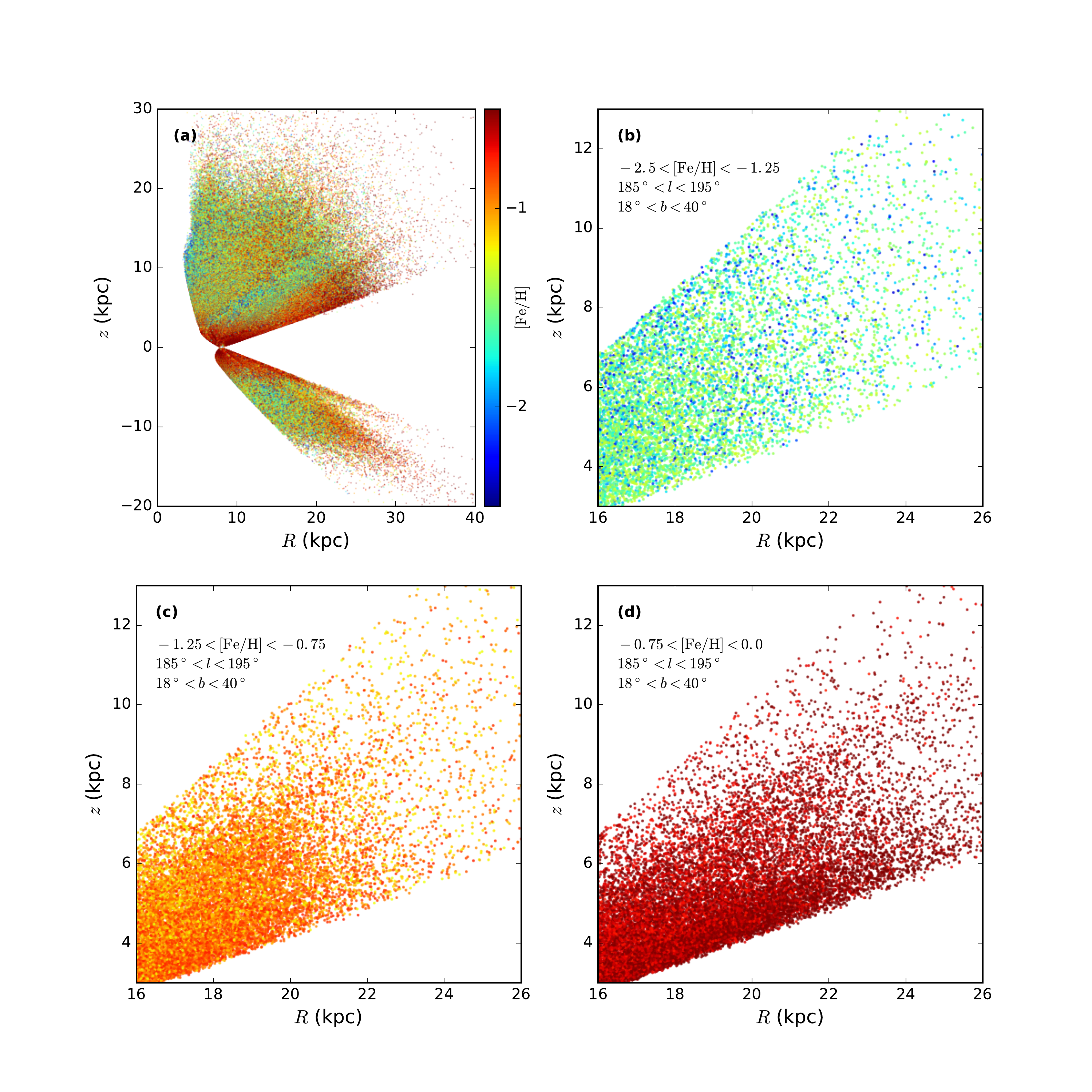}
\end{center}
\caption{(a) Metallicity distribution in the Galactic windows surveyed to date by CFIS-u. A large number of high-metallicity stars are seen at large $R$ and $z$, appearing to fan-out above the disk. Panels b--d concentrate on the region in the longitude range $185^\circ<\ell<195^\circ$, where the survey currently probes the Galactic Anticenter. It is clear that the metal-poor stars have a very different distribution to the intermediate (c) and metal-rich populations (d), as they possess a smoother spatial distribution. A ``boxy" structure extending to $z \sim 8\kpc$ is present both in (c) and (d), and additionally we detect a strong metal-rich component closer to the Galactic plane, extending out to $R\sim 25\kpc$.}
\label{fig:metallicity_map}
\end{figure*}

\section{The Galactic Anticenter}
\label{sec:Anticenter}

As a first example of what will be possible with CFIS-u, we will examine the populations towards the Galactic Anticenter, making use of the CFIS photometric metallicity determinations. In Paper~II, we describe in detail the procedure we follow to derive ${\rm [Fe/H]}$. Briefly, this consists of using an empirically-determined three-dimensional Legendre polynomial to interpolate an ${\rm [Fe/H]}$ value from a star's ${(u-g)_0}$, ${(g-r)_0}$ and ${(g-i)_0}$ colors (we use ``Method 2" from Paper~II). The resulting metallicity is then supplied to the I08 photometric parallax calibration (Equations 1--3 in Paper~II), together with a de-reddened $r$-band magnitude to obtain the star's distance. The ${g,r,i}$ magnitudes that we use here to probe the Galactic Anticenter are combined SDSS and PS1 magnitudes on the SDSS photometric system, with the color selection and quality criteria of the ``Wide cut," as described in Paper~II.

In Figure~\ref{fig:metallicity_map} we display a metallicity map of our Galaxy in the $R$--$z$ plane, as viewed by CFIS-u. A striking aspect of this map is the strong metal-rich population in the outer Galaxy at distances $15\kpc \la R \la 25\kpc$. This structure, named the Galactic Anticenter Stellar Structure (GASS), was discovered over a decade ago in SDSS counts towards the Anticenter direction \citep{Newberg:2002dy}, and has been found to encircle a large swath of the Milky Way, approximately parallel to the disk over at least $100\deg$ \citep{2003MNRAS.340L..21I,2004ApJ...615..732R,2006MNRAS.367L..69M,2007ApJ...668L.123M,Conn:2008bq,Sharma:2010em,2014ApJ...787...19M}. \citet{2014ApJ...791....9S} and \citet{2016ApJ...825..140M} show recent panoramic views of this structure. Early spectroscopic follow-up indicated that the structure is composed of metal rich stars with ${\rm [Fe/H] = -0.4 \pm 0.3}$, on roughly circular orbits, with relatively cold kinematics $\sigma_v=20\pm 4\kms$ \citep{2003ApJ...594L.119C,2003ApJ...594L.115R}. The stars that make up the structure are not young: the so-called Triangulum-Andromeda overdensity (located between 15 and $20\kpc$ along the line of sight towards M31), that appears to be related to the GASS, has been measured to have an age of $6$--$10\Gyr$ \citep{Sheffield:2014bf}.

Since the structure is located on the edge of the disk, it appears natural to suppose that it may be a warp or feature of that Galactic component, similar to the various overdensities that have been detected in the outskirts of the disk of the Andromeda galaxy \citep{Ferguson:2002is,2005ApJ...634..287I}. However, the stream-like aspect of the GASS would be considered proof of an accretion origin if it inhabited almost any other orbital plane. Indeed, \citet{2005ApJ...626..128P} considered this possibility, presenting an $N$-body model for the GASS in which it forms from stars that are tidally disrupted from an accreted dwarf galaxy, as this undergoes dynamical friction and is assimilated into the outer disk of the Milky Way. The initial orbital mis-alignment of the satellite with respect to the disk then naturally leads to a stream-like structure that snakes in and out of the disk, similar to the behavior of the observed overdensity. Further support for the accreted dwarf scenario of formation of the GASS was found in the chemical abundances derived from high-resolution spectra, which show an abundance pattern similar to what is seen in dwarf spheroidal galaxies and unlike that of the disk \citep{Chou:2010id}. This evidence is not conclusive, however, since star formation in the low density environments in the outer disks of galaxies will have different gas retention and chemical evolution compared to the inner disk regions \citep{Barnes:2012bg}.

Recent studies have explored again the Galactic origin of the GASS, suggesting that it is due to material that has been kicked out of the Galactic disk \citep{Xu:2015ir,PriceWhelan:2015kn}, possibly due to the reaction of the close passage of an interloper \citep{2008ApJ...688..254K}, and/or the action of subsequent dynamical waves traveling through the disk.\footnote{The reader may find it useful to refer to figure~8 of \citet{Li:2017jg}, which sketches a possible configuration for the oscillatory behavior of the outer disk.} This latter possibility is particularly interesting in the light of the discovery of vertical waves in stellar density and mean vertical stellar motions in the Solar vicinity \citep{2012ApJ...750L..41W}. Subsequently, \citet{2014MNRAS.440.1971W} analyzed a simulated Galactic disk interacting with multiple satellites, in which the inner Galactic disk was shown to be prone to a strong bar instability and affected by vertical breathing modes,\footnote{A vertical breathing mode is a density perturbation with even parity with respect to the Galactic plane (i.e. no North-South asymmetry in density) and a vertical velocity field perturbation with odd parity. It is thus a vertical ``rarefaction-compression'' wave.} whilst the outer disk was shown to be affected by bending modes.\footnote{A vertical bending mode has odd parity in density with respect to the plane, and even parity in the vertical velocity field. It is thus a vertical ``corrugation" wave.} Such a vertical bending mode could precisely be responsible for the GASS. Interestingly, \citet{2015MNRAS.452..747M,2016MNRAS.457.2569M,2016MNRAS.461.3835M} managed to explain qualitatively the properties of the inner disk breathing modes of the \citet{2014MNRAS.440.1971W} simulation from the effects of its bar and spirals (see also \citealt{Faure:2014bn}). A vertical breathing mode is actually observed locally in the kinematics of red clump stars out to $|z| \sim 2\kpc$ \citep{Williams:2013dq}, but with a rather high vertical velocity amplitude compared to the predictions from spiral density waves in \citet{2016MNRAS.457.2569M,2016MNRAS.461.3835M}, a discrepancy potentially due to the transient nature of spirals, which could themselves be induced by interactions with satellites. Thus, the inner disk breathing mode, as seen by \citet{Williams:2013dq}, could be related to satellite-induced non-axisymmetries of the disk, whilst the GASS could be the signature of the bending mode induced by those same satellite interactions. In this picture, the Solar vicinity would then be in the transition zone between the inner breathing Galactic disk and the outer bending Galactic disk. The consequences of this emerging complex picture on our future dynamical modeling of the Milky Way in the \Gaia/CFIS era are profound. It was for instance recently shown that local dynamical determinations of the dark matter density could be off by $\sim 25$\% if one wrongly assumes that the Solar neighborhood is in equilibrium \citep{2017MNRAS.464.3775B}. 

\begin{figure}
\begin{center}
\includegraphics[angle=0, viewport= 10 15 425 420, clip, width=\hsize]{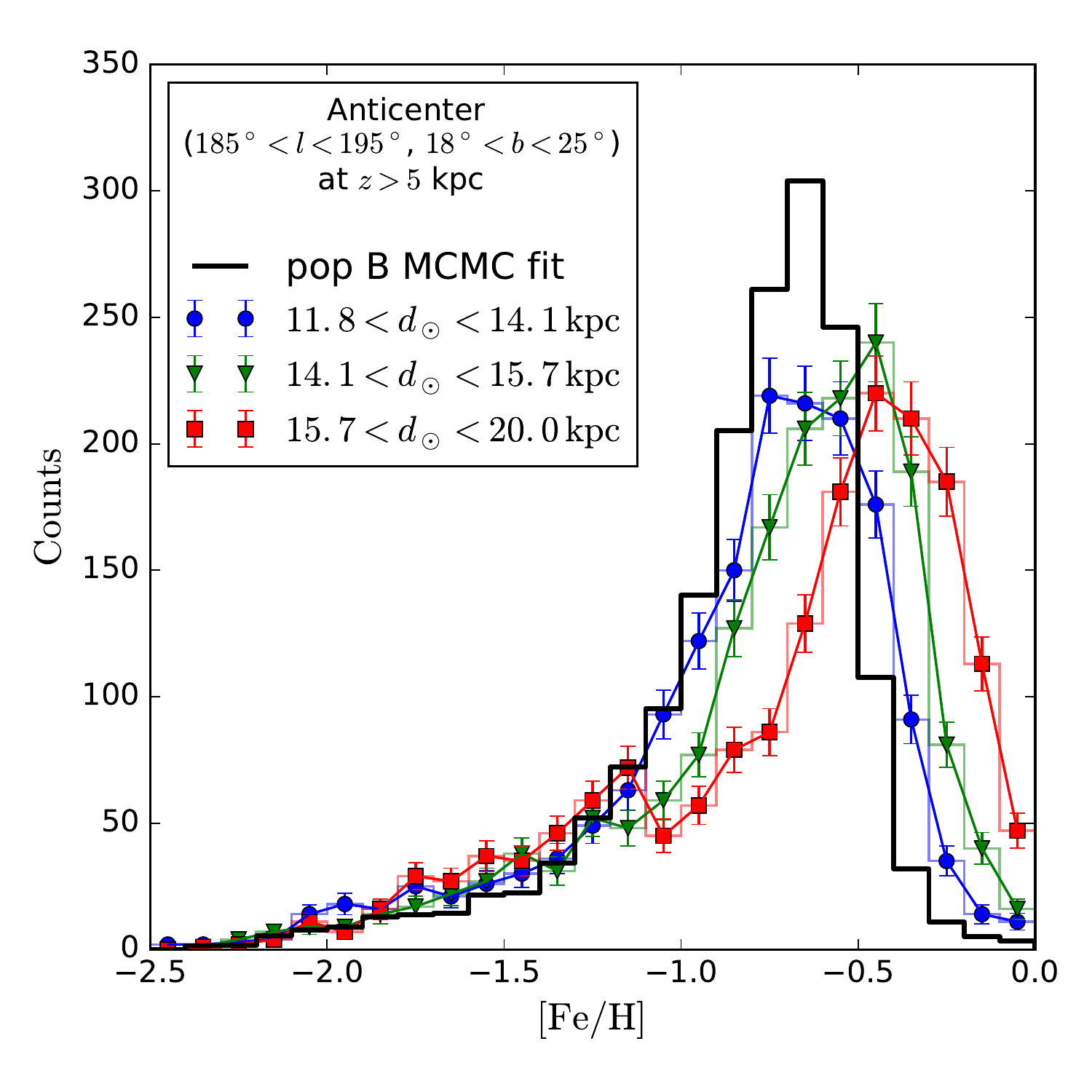}
\end{center}
\caption{Metallicity distributions at large distance above the Galactic plane ($z>5\kpc$) in the CFIS-u field closest to the Galactic Anticenter. The colored points show three different distance selections (the limits are chosen to have an approximately equal number of counts in each sample). Up to a Heliocentric distance of $15.9\kpc$ the distributions can be seen to be similar to the local Solar neighborhood thick disk component (black histogram, dubbed ``pop B'' in Paper~II) fitted by the Monte Carlo procedure shown in figure~12a of Paper~II to data towards the North Galactic Cap. The outermost sample appears more metal-rich.}
\label{fig:metallicity_distribution_GASS}
\end{figure}

The high-metallicity sub-component seen in the CFIS-u map of Figure~\ref{fig:metallicity_map}a is present at high vertical distance $z$ over the entire Galactic anticenter direction that CFIS-u currently probes. Figures~\ref{fig:metallicity_map}b--d show the distribution of three metallicity slices towards the Galactic anticenter, selected from the spatial box $185\deg < l < 195\deg$, $18\deg< b < 40\deg$. The flattened feature is present predominantly in metal-rich stars (${\rm [Fe/H]>-0.75}$), where it forms a continuous extended structure out to $R\sim 25\kpc$.

\begin{figure}
\begin{center}
\includegraphics[angle=0, viewport= 10 15 425 420, clip, width=\hsize]{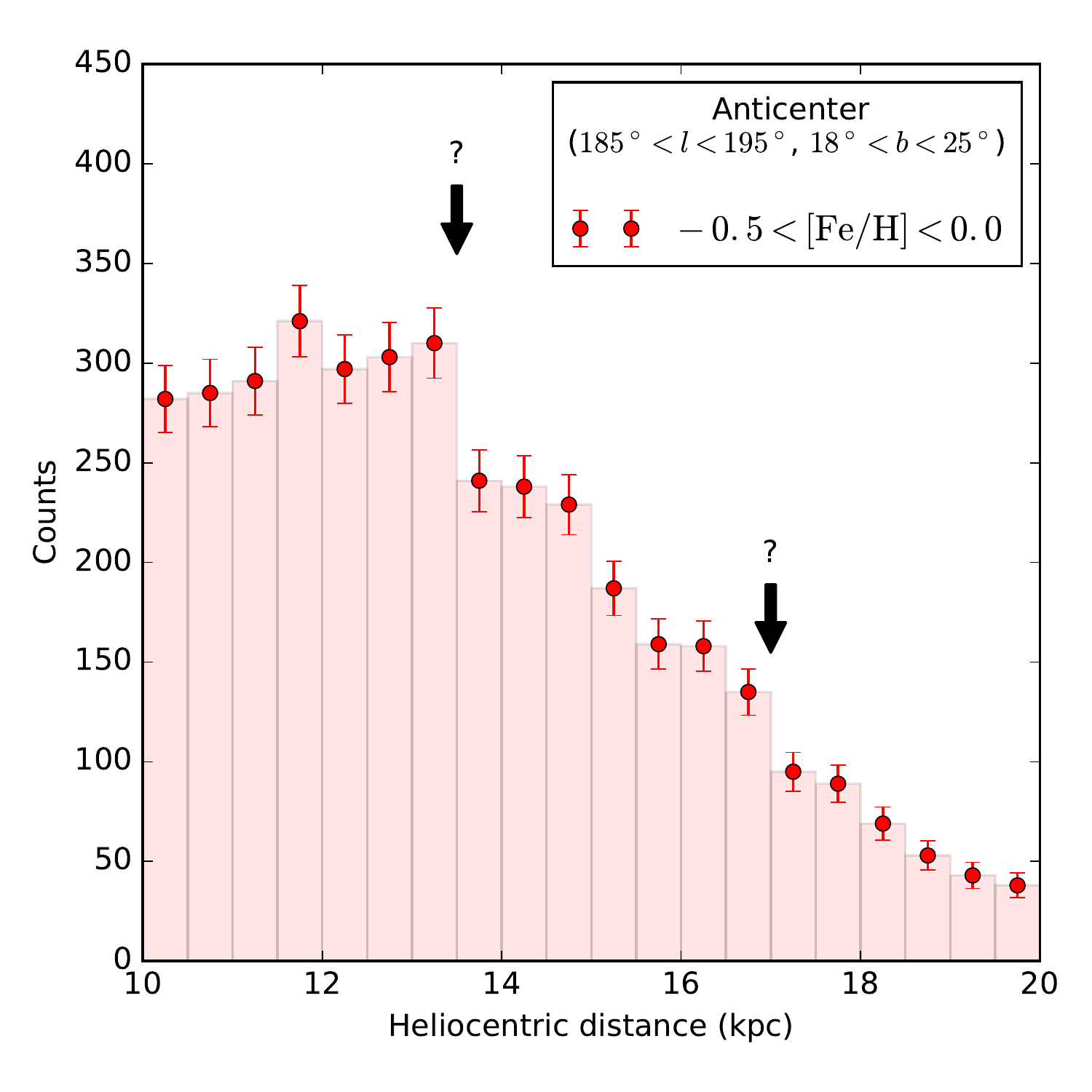}
\end{center}
\caption{The distribution along the line of sight of metal-rich stars towards the Galactic Anticenter (same sky area as in Figure~\ref{fig:metallicity_distribution_GASS}). The line of sight distribution is approximately smooth, although there may be some discontinuities at the locations marked by the arrows. These will be investigated further when the survey covers a greater area of the Galactic Anticenter region.}
\label{fig:GASS_los}
\end{figure}

The metallicity distribution in this Anticenter field at $z>5\kpc$ is shown in Figure~\ref{fig:metallicity_distribution_GASS}. The limit of $z=5\kpc$ was chosen as this corresponds to the distance in figures~12 and 13 of Paper~II where the metal-poor halo population (the ``pop C'' component of the metallicity-distance decomposition) becomes dominant. However, it can be seen from this diagram that that is clearly not the case towards the Galactic Anticenter, since the metal-rich population is dominant far above the plane.

It is interesting to compare these properties with what we have learned in recent years about the outer disk of the Milky Way, in particular from APOGEE. The results from APOGEE \citep[][their figure 4]{Hayden:2015es} show that the outer disk at a distance of $11$--$15\kpc$ (the farthest distance range probed by that survey) has very simple chemical patterns. The distributions of metallicity and alpha abundance show peaks at ${\rm [Fe/H]\sim -0.4}$~dex and ${\rm [\alpha/Fe]\sim0.07}$ dex, with a limited dispersion in both parameters of $0.17$--$0.19$~dex, while the spread in metallicity goes from $-0.6$ to $0.0$~dex. The stars in the high-metallicity component in the metallicity-distance decompositions presented in Paper~II peak at the same metallicity, and most have ${\rm [Fe/H]>-0.6}$ dex (figure 13 of Paper~II). Its metallicity distribution function (MDF) is remarkably narrow given the volume sampled and the uncertainties of the photometric metallicities, and is very similar to the MDF sampled by APOGEE at $13<R<15\kpc$. 

In Figure \ref{fig:metallicity_distribution_GASS} (which shows stars at $z>5\kpc$), the nearer component (blue line) has a significantly lower peak metallicity (around $-0.7$~dex) than the outer disk in APOGEE. Indeed, its MDF is similar to the intermediate metallicity component identified in the metallicity-distance decomposition in Paper~II (black line), which we attribute to the Solar neighborhood thick disk. As we probe out further in distance, however (green and red lines), the distributions become substantially more metal rich. Given that these more distant samples also contain a contribution from the intermediate component as well as halo stars, the metal-rich remainder population probably peaks at slightly higher values that what we see here. Taking into account the small but significant negative radial metallicity gradient observed in the outer disk, the mean metallicity of the disk population probably decreases between $R=16$ and $26\kpc$ compared to what it is at $R<15\kpc$. The outer disk, as it is known from APOGEE within $R<15\kpc$, therefore appears compatible with these CFIS-u data, giving support to the hypothesis that the GASS is probably the result of the outer disk being pushed out of the Galactic plane. 

\begin{figure}
\includegraphics[angle=0, viewport= 25 205 540 620, clip, width=\hsize]{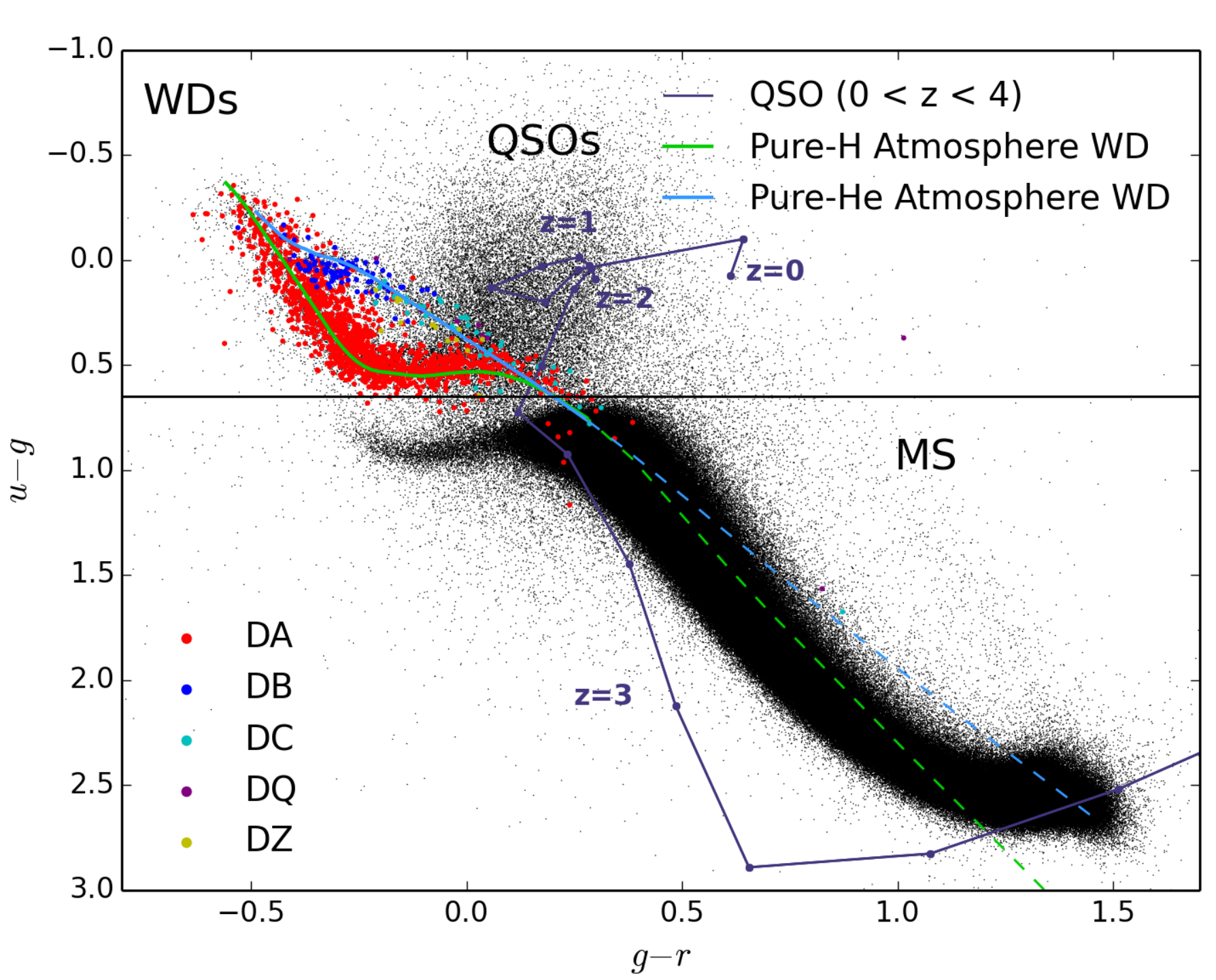}
\caption{Color-color diagram showing the location of spectroscopically confirmed white dwarfs from \cite{2013ApJS..204....5K} sorted by type.  Model QSO tracks from \cite{2009AJ....137.3761B} are shown in purple and model white dwarfs with pure hydrogen or pure helium atmospheres from \cite{2006AJ....132.1221H} are shown in green and blue respectively, transitioning from solid to dashed in the region of the main sequence (for clarity). }
\label{fig:WD}
\end{figure}

Although the MDF of Figure~\ref{fig:metallicity_distribution_GASS} peaks at a metallicity typical of the thick disk, we exclude the possibility that the metal-rich subcomponent is the thick disk itself -- defined from its chemical properties as an alpha-rich population, implying a number of specific characteristics \citep[see][]{2013A&A...560A.109H}. It has been shown that the short scale-length of this population essentially confines the thick disk to the inner $10\kpc$ of the Milky Way \citep{2011ApJ...735L..46B,2012ApJ...752...51C,2012ApJ...753..148B}. In fact, the outer disk (as least as far out as $R=15\kpc$) is exclusively populated by a low-alpha population, as shown by APOGEE \citep{Hayden:2015es}.  

The distance distribution of the metal-rich stars towards the Galactic Anticenter is shown in Figure~\ref{fig:GASS_los}, where we plot stars with ${\rm -0.5 < [Fe/H] < 0}$ in the same window of sky as before. The distribution is approximately smooth, although there are hints of discontinuities at the locations marked with arrows. We will attempt to verify whether these features exist in adjacent regions of the sky once the CFIS-u survey covers the anticenter more completely.

The relative smoothness of the stellar distribution along this line of sight to the Galactic anticenter, as well as the chemical similarity with the 
outer disk (Figure~\ref{fig:metallicity_distribution_GASS}) suggests that the GASS resembles a warped disk of old stars. It will therefore be very interesting to obtain further $u$-band data closer to the Galactic plane so as to probe the asymmetry of the outer disk as a function of distance.

\begin{figure*}
\begin{center}
\includegraphics[angle=0, viewport= 70 1 1170 330, clip, width=\hsize]{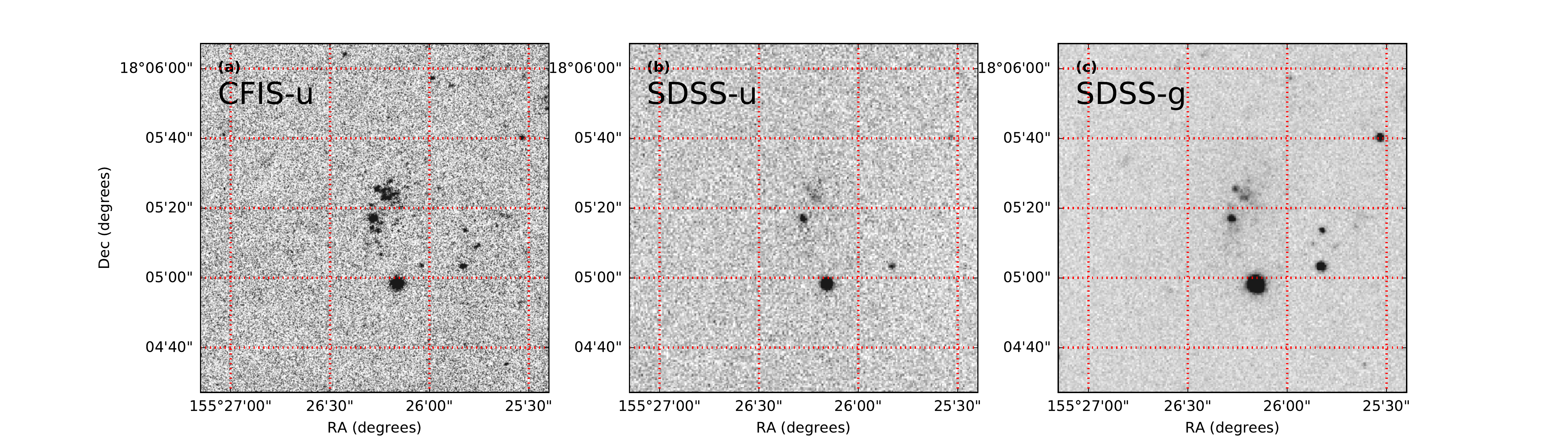}
\end{center}
\caption{The Leo-P star-forming dwarf galaxy in CFIS-u (a) and in the SDSS $u$-band (b) and $g$-band (c). The structure stands out as a clear overdensity of $\sim 20$ point sources in CFIS-u.}
\label{fig:Leo P}
\end{figure*}

\section{White dwarfs and horizontal branch stars}
\label{sec:WDs}

White dwarfs are the remnants of some of the oldest stars in the Galaxy and have provided vital information regarding the formation, evolution, and structure of the Milky Way. Recent large surveys, such as the SDSS, have increased the number of spectroscopically confirmed white dwarfs by more than an order of magnitude, allowing for large statistical analyses of the Galactic population. Much of the focus has been on the coolest and oldest white dwarfs, since they can provide age constraints for the Galactic components \citep[see, e.g.,][]{1998ApJ...497..294L,2012Natur.486...90K,2015MNRAS.449.3966G}. On the other hand, studies of the hot, and hence young, white dwarf population focus on constraining the initial-final mass relationship --- an important input for stellar population synthesis codes \citep[see, e.g.,][]{2011MNRAS.411.2770B}.

The combination of wide sky coverage, $u$-band depth, and the ability to cross-match with other deep photometric surveys (e.g., SDSS, PS1, or the Next Generation Virgo Survey -- \citealt{2012ApJS..200....4F}) makes CFIS-u ideally suited to select a large number of white dwarfs. Figure \ref{fig:WD} shows a ${(u - g)}$, ${(g - r)}$ color-color diagram of all CFIS-u sources with $\delta$u, $\delta$g, $\delta$r $<$ 0.05 mag and u $>$ 17 mag. Spectroscopically confirmed white dwarfs identified by \cite{2013ApJS..204....5K}, sorted by type, are shown as colored points. The theoretical sequence for white dwarfs with a mass of $M = 0.6 \msun$ and atmospheres of pure hydrogen or pure helium (computed from \citealt{2006AJ....132.1221H} model atmospheres\footnote{\tt{http://www.astro.umontreal.ca/\~{ }bergeron/CoolingModels}}) are shown in green and purple respectively. Figure \ref{fig:WD} shows the clear separation between the white dwarf cooling sequence, QSOs, and main-sequence stars.

The positions and $u$-band magnitudes will also be important for the cool end of the white dwarf luminosity function. White dwarf colors turn blueward at low temperatures as a result of collision induced opacity from molecular hydrogen, and hence deep $u$-band photometry will aid in the detection of these cool objects. However, cool white dwarfs are difficult to detect as they are faint and have similar colors to main sequence stars. Previous attempts to disentangle the coolest white dwarfs from other populations have relied on their proper motions \citep[see, e.g.,][]{2006AJ....131..571H,2006AJ....131..582K,2011MNRAS.417...93R}. Combining the astrometry from SDSS with that of CFIS will yield proper motions for sources with deep $u$-band photometry. The combination of $u$-band magnitude and proper motions will allow for a comprehensive selection of cool white dwarfs for follow-up spectroscopy. 

Sub-dwarf O and B stars are the most common hot stars in the halo apart from white dwarfs, and CFIS-u should readily allow their photometric identification \citep{2014ApJS..215...24S}. The depth of CFIS-u is also sufficient to identify effectively all A-type stars, including horizontal branch stars, out to beyond the virial radius of the Galaxy (D$\sim$300kpc), which will allow us to study the global shape and substructure of the distant outer halo. This will provide targets for follow-up spectroscopy that will be used to constrain the mass distribution of the Milky Way at very large radius. While blue stragglers will contaminate the sample of horizontal branch stars, the former have higher surface gravity and can be distinguished from the latter with $ugr$ photometry \citep{2011MNRAS.416.2903D}.

\section{Star-formation on the edge of the Local Group}
\label{sec:star-formation}

Another very interesting issue that can be investigated with a survey such as CFIS-u is to quantify the prevalence of star-forming dwarf galaxies in the environment surrounding the Local Group. These objects may trace the recent accretion of dark and baryonic matter into our environment and can potentially tell us about the suppression of star-formation in isolated low-mass haloes \citep{Read:2016ww}.
 
The prototypical object of this class is Leo-P, discovered by \citet{2013AJ....146...15G} in a follow-up of compact high-velocity \ion{H}{1} clouds detected by the ALFALFA survey \citep{2005AJ....130.2598G}. Located at a distance of $(1.62\pm0.15) \mpc$ \citep{2015ApJ...812..158M}, this small galaxy appears to be an unquenched analogue to the dwarf spheroidals that are observed in abundance around the Milky Way and Andromeda.

Whether such objects are numerous or not remains an open question, as they have been very difficult to detect. In the SDSS, Leo-P appears as a very low-surface brightness structure with 2--3 faint point sources superimposed (Figure~\ref{fig:Leo P}b and \ref{fig:Leo P}c). In contrast, the greater depth and much better image quality of CFIS-u allows us to resolve the structure into around $20$ point sources (Figure~\ref{fig:Leo P}a). 

This demonstrates the feasibility of detecting similar galaxies simply by searching for localised overdensities of point sources in our $u$-band maps.

\begin{figure*}
\begin{center}
\includegraphics[angle=0, viewport= 134 1 970 322, clip, width=\hsize]{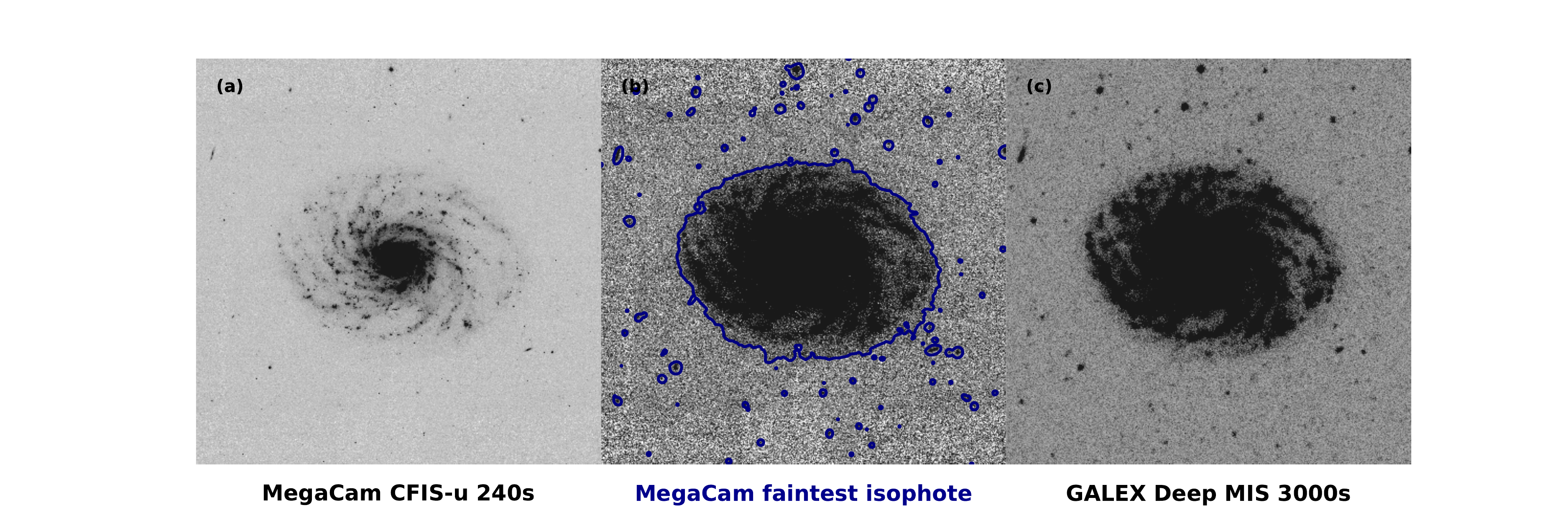}
\end{center}
\caption{Example of the LSB performance of CFIS-u. The target here is a $11\mcnp\times11\mcnp$ region around the nearby spiral galaxy NGC~3486. The MegaCam image (a) is reproduced in (b) with a stretch that better reveals the low surface brightness emission. The contours show the faintest isophotes (derived after convolving the image with a $1\scnp$ Gaussian kernel). One can appreciate that CFIS-u reaches a similar depth to low-surface brightness emission as the GALEX Medium Imaging Survey (c), while possessing much better spatial resolution: the radial extent of the galaxy is almost identical to that seen in the GALEX image, and also almost all of the faint background sources in GALEX are present in the CFHT contour map.}
\label{fig:LSB}
\end{figure*}

\section{The \MakeLowercase{{\it u}}-band low surface brightness universe}
\label{sec:LSB}

We have investigated the use of the CFIS-u exposures for uncovering low-surface brightness structures in ultraviolet light. This is potentially a very interesting application of the CFIS-u data, since it may allow us to uncover and quantify low-levels of star-formation in the extreme outskirts of galaxies, opening up the possibility
of studying star-formation as a function of environment in a large volume of the local Universe. 

To this end we adapted the Elixir-LSB software, which was initially developed to process the Next Generation Virgo Survey (NGVS, \citealt{2012ApJS..200....4F}), to the particular case of the CFIS-u signal properties and observing strategy. A sky model is constructed using a sliding window of a minimum of nine sequential frames in time: with a new frame captured every four minutes, the intrinsic variation of the sky in time can be frozen. The NGVS showed that a window of up to one hour long exhibits stable behaviour in the $u$-band for a moonless sky, reaching after Elixir-LSB processing $29 \, {\rm mag \, arcsec^{-2}}$ in direct detection performance ($32 \, {\rm mag \, arcsec^{-2}}$ when integrating a galaxy profile). 

With CFIS-u, we have the luxury of having many short exposures and find that reducing the window to $\sim 30$ minutes, while still having enough frames to reject astronomical sources from the background, gives the best results. To restore the true sky signal in an image, we associate with each image a list of input frames captured within that time window, out of which a map correction frame is extracted. For long periods of CFIS-u observations, the list consists in the four preceding and four following frames (and skewed lists for the very end and beginning of the CFIS-u set on that sequence of the night). For each list of nine images, we derive the precise sky level for each frame and perform a median combination after scaling each image to a common median background value, and after having also smoothed each frame with a $2\mcnp$ wide Gaussian kernel. This smoothing ensures the sigma-clipping algorithm properly rejects faint signals above the background, such as the extended faint features we aim to detect or optics reflection halos and other artifacts. This median background map is then precisely scaled back to the background of the central image of the set it is associated with, to enable a precise background subtraction, restoring the true sky signal. 

Using a ${\rm (min-max)/max}$ metric on the corrected image background (searching for the faintest true signal above the noisy background, i.e., detecting by eye the faintest extended feature), the CFIS-u data show a limit at the 1\% level (to be compared to 0.2\% for the NGVS $u$-band), or equivalently, this indicates that we detect features that are in contrast 5 magnitudes fainter than the sky background ($\sim 28 \, {\rm mag \,arcsec^{-2}}$). The performance is not as good as the NGVS: this is due to the noise properties of the signal, which is not in the pure photon-noise dominated regime: the typical CFIS-u 80~s exposure has a background level of $\sim 30$ electrons, giving a photon noise of just $\sim 6$ electrons, which is comparable to the CCD readout noise. This limits the Elixir-LSB performance, since the true background level remains biased. The pipeline will be updated to include a more accurate noise model when running the sigma-clipping rejection. 

We selected NGC 3486, a nearby spiral galaxy, to test the procedure described above. The result is shown in Figure~\ref{fig:LSB}, where the CFIS-u image is compared to the identical region from the GALEX Medium Imaging Survey (MIS, \citealt{2005ApJ...619L...1M}). Compared to the 3000~s GALEX MIS image, the 240~s CFIS-u LSB-stack appears competitive in depth, and has much better spatial resolution. We will explore the properties of the low surface-brightness components of nearby galaxies in forthcoming contributions in this series.

\section{Discussion and conclusions}
\label{sec:Conclusions}

We have presented the $u$-band component of the new Canada-France Imaging Survey, a community effort to obtain $u$- and $r$-band photometry of the northern sky that will be used for a host of stand-alone science studies as well as being part of the requirements of the \Euclid\ mission to measure photometric redshifts of galaxies at cosmological distances. Our plan is to release CFIS calibrated images and photometric catalogues to the international community in early 2021, one year after the end of the observing campaign. Here we have focussed on some highlights of the Local Universe science that will be possible with the $u$-band, thanks to the depth and excellent image quality of the survey, using the $\sim$2,900$\, {\rm deg^2}$ already observed, approximately one third of the final $u$-band survey. 

CFIS-u currently probes only a relatively small region towards the Galactic Anticenter, but the data in-hand already allow us to measure the chemical properties of the outskirts of the Galactic disk in this direction, and probe the variation of the structures along the line of sight. The metallicity distribution of the metal-rich component in the outer disk is remarkably narrow, given the volume sampled, with a dispersion only slightly larger than the uncertainties in the metallicities, pointing to a possibly very homogeneous population. Finally, the resemblance of the metallicity distribution in the outer disk at high extraplanar distance ($z>5\kpc$) with the disk sampled by APOGEE suggests that the metal-rich stars found towards the Galactic Anticenter are most likely simply the outer disk population that has been kinematically heated or warped by minor accretion events.

Upcoming contributions in this series will investigate white dwarf populations in our Galaxy, as well as attempting to identify blue straggler and blue horizontal branch stars to probe the outermost reaches of the halo of the Milky Way, and presenting a new metallicity calibration for giant branch stars. We expect the CFIS-u metallicities and distances (presented in detail in Paper~II) to be particularly useful when they are joined with faint sources from \Gaia, since they will allow us to convert proper motions to transverse velocities, hence placing the populations in distance, and allowing population discrimination; all of this will greatly increase the power of the \Gaia\ dataset for halo science. The $u$-band sensitivity will also be used to attempt to quantify the prevalence of nearby star-forming dwarf galaxies, and we will investigate the low-surface brightness Universe in the u band.

\acknowledgments

We thank the staff of the Canada-France-Hawaii Telescope for taking the CFIS data and their extraordinary support throughout the project. We are especially indebted to Todd Burdulis for the care and dedication given to planning and observing this survey. RAI and NFM gratefully acknowledge support from a ``Programme National Cosmologie et Galaxies'' grant.

This work is based on data obtained as part of the Canada-France Imaging Survey, a CFHT large program of the National Research Council of Canada and the French Centre National de la Recherche Scientifique. Based on observations obtained with MegaPrime/MegaCam, a joint project of CFHT and CEA Saclay, at the Canada-France-Hawaii Telescope (CFHT) which is operated by the National Research Council (NRC) of Canada, the Institut National des Science de l'Univers (INSU) of the Centre National de la Recherche Scientifique (CNRS) of France, and the University of Hawaii. This research used the facilities of the Canadian Astronomy Data Centre operated by the National Research Council of Canada with the support of the Canadian Space Agency.

Funding for the Sloan Digital Sky Survey IV has been provided by the Alfred P. Sloan Foundation, the U.S. Department of Energy Office of Science, and the Participating Institutions. SDSS-IV acknowledges support and resources from the Center for High-Performance Computing at the University of Utah. The SDSS web site is \url{www.sdss.org}. SDSS-IV is managed by the Astrophysical Research Consortium for the Participating Institutions of the SDSS Collaboration including the Brazilian Participation Group, the Carnegie Institution for Science, Carnegie Mellon University, the Chilean Participation Group, the French Participation Group, Harvard-Smithsonian Center for Astrophysics, Instituto de Astrof\'isica de Canarias, The Johns Hopkins University, Kavli Institute for the Physics and Mathematics of the Universe (IPMU) / University of Tokyo, Lawrence Berkeley National Laboratory, Leibniz Institut f\"ur Astrophysik Potsdam (AIP), Max-Planck-Institut f\"ur Astronomie (MPIA Heidelberg), Max-Planck-Institut f\"ur Astrophysik (MPA Garching), Max-Planck-Institut f\"ur Extraterrestrische Physik (MPE), National Astronomical Observatories of China, New Mexico State University, New York University, University of Notre Dame, 
Observat\'ario Nacional / MCTI, The Ohio State University, Pennsylvania State University, Shanghai Astronomical Observatory, 
United Kingdom Participation Group, Universidad Nacional Aut\'onoma de M\'exico, University of Arizona, University of Colorado Boulder, University of Oxford, University of Portsmouth, University of Utah, University of Virginia, University of Washington, University of Wisconsin, Vanderbilt University, and Yale University.

The Pan-STARRS1 Surveys (PS1) have been made possible through contributions of the Institute for Astronomy, the University of Hawaii, the Pan-STARRS Project Office, the Max-Planck Society and its participating institutes, the Max Planck Institute for Astronomy, Heidelberg and the Max Planck Institute for Extraterrestrial Physics, Garching, The Johns Hopkins University, Durham University, the University of Edinburgh, Queen's University Belfast, the Harvard-Smithsonian Center for Astrophysics, the Las Cumbres Observatory Global Telescope Network Incorporated, the National Central University of Taiwan, the Space Telescope Science Institute, the National Aeronautics and Space Administration under Grant No. NNX08AR22G issued through the Planetary Science Division of the NASA Science Mission Directorate, the National Science Foundation under Grant No. AST-1238877, the University of Maryland, and Eotvos Lorand University (ELTE).

\bibliography{ms}
\bibliographystyle{apj}

\end{document}